\documentclass[secnumarabic,amssymb, superscriptaddress, nobibnotes, aps,prl,preprint]{revtex4-1}
\usepackage{amsmath}
\usepackage{graphicx,nicefrac}
\usepackage{mathtools}
\usepackage{xcolor}

\usepackage{harpoon}
\newcommand{\vect}[1]{\overrightharp{\ensuremath{#1}}}

\newcommand{\Fex}[1]{Fe$_{#1}$NbS$_2$}
\newcommand{\um}{\mu m}

\begin{document}
\title{Long-range, Non-local Switching of Spin Textures in a Frustrated Antiferromagnet}

\newcommand{\UCB}{Department of Physics, University of California, Berkeley, CA 94720, USA}
\newcommand{\LBL}{Materials Sciences Division, Lawrence Berkeley National Laboratory, Berkeley, California, 94720, USA}

\newcommand{\UCBC}{Department of Chemistry, University of California, Berkeley, California 94720, USA}
 
\newcommand{\NHMFL}{National High Magnetic Field Laboratory, Tallahassee, Florida 32310, USA}

\newcommand{\UCBM}{Department of Materials Science and Engineering, University of California, Berkeley, California 94720, USA}

\author{Shannon C. Haley $^\ast$}
\affiliation{\UCB}
\affiliation{\LBL}

\author{Eran Maniv}
\affiliation{\UCB}
\affiliation{\LBL}

\author{Shan Wu}
\affiliation{\UCB}

\author{Tessa Cookmeyer}
\affiliation{\UCB}
\affiliation{\LBL}

\author{Susana Torres-Londono}
\affiliation{\UCB}

\author{Meera Aravinth}
\affiliation{\UCB}

\author{Nikola Maksimovic}
\affiliation{\UCB}
\affiliation{\LBL}

\author{Joel Moore}
\affiliation{\UCB}
\affiliation{\LBL}

\author{Robert J. Birgeneau}
\affiliation{\UCB}

\author{James G. Analytis $^\ast$}
\affiliation{\UCB}
\affiliation{\LBL}

\date{\today}

\begin{abstract}
Antiferromagnetic spintronics is an emerging area of quantum technologies that leverage the coupling between spin and orbital degrees of freedom in exotic materials. Spin-orbit interactions allow spin or angular momentum to be injected via electrical stimuli to manipulate the spin texture of a material, enabling the storage of information and energy. In general, the physical process is intrinsically local: spin is carried by an electrical current, imparted into the magnetic system, and the spin texture then rotates. The collective excitations of complex spin textures have rarely been utilized in this context, even though they can in principle transport spin over much longer distances, using much lower power. In this study, we show that spin information can be transported and stored non-locally in the material Fe$_{x}$NbS$_2$. We propose that collective modes leverage the strong magnetoelastic coupling in the system to achieve this, revealing a novel way to store spin information in complex magnetic systems.

\end{abstract}

\maketitle

\normalsize{$^\ast$Corresponding authors. Email: shannon\_haley@berkeley.edu and analytis@berkeley.edu}

\section{Introduction}

The semiconductor devices behind modern computers are rapidly approaching the physical limits of charge-based electronics, spurring research into novel materials that can enable `spintronic' technologies that leverage the spin as well as the charge of an electron. Magnonics is an emerging subfield whereby the collective excitations of the magnetically ordered system, known as magnons, can be electrically stimulated.\cite{chumak_magnon_2015} Such materials have unique advantages because the length scale over which spin is coherently transported without loss can be very large, in contrast to flowing electrons whose spin decay is generally shorter.\cite{short_afm_metal} In addition to spin coherence, there is also the challenge of spin-based memory. It has been shown that some antiferromagnetic (AFM) materials can store spin information through the electrical manipulation of AFM domains, although such technologies are thought to use spin polarized electrical currents that, on general grounds, are intrinsically local in nature.\cite{wadley_electrical_2016,bodnar_2018}  However, the role of strain in these phenomena has risen in prominence recently, as the electrical manipulation of certain insulating antiferromagnets has been shown to be driven by a combination of strain and thermal effects \cite{kent_AFM,afm_thermomagnetoelastic} - and with it, there is a growing understanding of the possibilities for switching broader classes of antiferromagnets and for longer-range manipulations.  In this study, we show that a class of switchable, metallic antiferromagnets \Fex{x}, exhibits the ability to manipulate spin information `non-locally' -- namely, tens of microns away from the electrical stimulus. This is orders of magnitude further than the spin diffusion length of typical metallic antiferromagnets\cite{lebrun_long-distance_2020,han_AFMmagnons,nonlocal_spin_superfluid,lebrun_transport2018, short_afm_metal}. We propose a picture that leverages two long range effects: collective excitations to carry spin and strong magnetoelastic coupling to allow complex domain structures to propagate over large distances. 

\begin{figure}
    \centering
    \includegraphics[width=\linewidth]{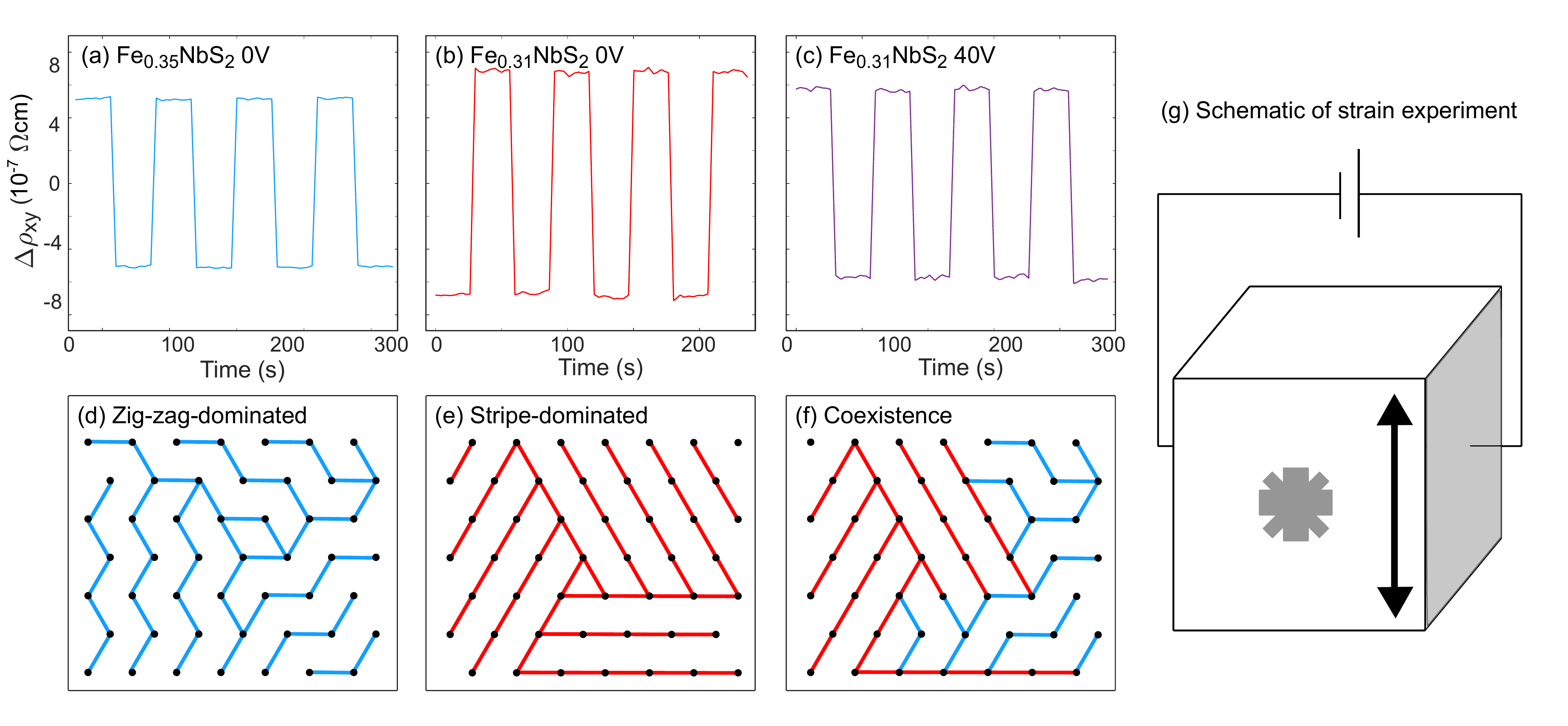}
    \caption{Switching behavior under applied uniaxial strain. (a)-(c) The change in transverse resistivity while $10ms$ DC current pulses of about $5\times10^{-4}A/cm^2$ are applied in alternating directions. (a) Measurement for a device made of \Fex{0.35} with no applied strain. (b) Measurement for a device made of made of \Fex{0.31} with no applied strain. (c) Measurement for a device made of \Fex{0.31} following cooling from room temperature with applied strain corresponding to $40V$ on the piezoelectric cube the device was mounted on. (b) and (c) were measured on the same device. Complete dataset with pulse current dependence is available in the supplement (Fig. \ref{fig:piezo_extended}). (d) Dominant spin texture in \Fex{0.35}. (e) Dominant spin texture in \Fex{0.31}. (f) Proposed spin texture in \Fex{0.31} cooled under strain. Proportion of zigzag phase is exaggerated. (g) Schematic of strain measurement. Voltage is applied between two electrodes around a cube of piezoelectric material, causing a directional expansion of the material which strains the device mounted on the cube.   }
    \label{fig:piezo}
\end{figure}





The compound \Fex{x} is an easy-axis antiferromagnet on a triangular sublattice which has been found to switch between distinct resistance states upon the application of DC current pulses along perpendicular directions \cite{nair_electrical_2020}. Importantly, it appears that collective dynamics of the magnetic spin texture play an important role with this directional switching, with very high tunability by compositional changes about $x\simeq \nicefrac{1}{3}$ \cite{maniv_antiferromagnetic_2021}. This switching behavior was originally ascribed to a $90\deg$ reorientation of the in-plane component of the N\'eel vector, imparted via spin-orbit torque from conduction electrons which become spin-polarized due to the Rashba-Edelstein Effect, which is allowed because of the broken inversion symmetry of the crystal lattice \cite{nair_electrical_2020, bodnar_2018, wadley_2016, zelezny}.  Further study found that the single-ion anisotropy in this system is strong enough to preclude the possibility of any significant in-plane moment, and so differing orientations of the N\'eel vector alone cannot be responsible for the anisotropic resistance we observe \cite{haley2020halfmagnetization}. Instead, recent work has shown that there are two nematic and nearly degenerate antiferromagnetic ground states in \Fex{x}, one in which aligned spins form stripes and one in which they form zig-zags \cite{wu2021highly}. The current pulse appears to rotate the principal nematic axis of the magnetic order. These orders compete, with the stripe order dominating at dilute compositions $x<\nicefrac{1}{3}$, and zig-zag at excess Fe composition $x>\nicefrac{1}{3}$ \cite{wu2021highly}. According to a recent DFT study, which explores the Fermi surface anisotropies of the domains of the respective phases, this likely explains the opposite switching responses in identical device geometries, as shown in Fig. \ref{fig:piezo}(a) and (b) \cite{weber2021}.  Consider the domains structures in Fig. \ref{fig:piezo}(d) -- (f). A given direction of switching pulse ${\bf \vect{j}}$ destabilizes domains whose principal axes are parallel to the applied current, so that a pulse in the [100] direction will strongly disfavor one specific stripy domain and one specific zig-zag domain \cite{weber2021,gao_relativistic_2014}. With respect to the principal axes, the conductivity tensor components $\sigma_{xx}>\sigma_{yy}$ for stripe domains and $\sigma_{xx}<\sigma_{yy}$ for zigzag domains, so when the current is applied along $45^\circ$, there are opposite switching responses in the off-diagonal resistivity \cite{weber2021}. At compositions where the order parameters are comparable in magnitude, one would expect the response to vanish -- and this is exactly what is observed at $x = 0.33$, where the amplitude of the switching response is suppressed and a change in the sign of the response is observed as a function of the pulse current density \cite{maniv_antiferromagnetic_2021}.

The two order parameters are known to have strong magneto-elastic coupling\cite{little2020} and so it is likely that strain can be used to tune the switching behavior. To demonstrate this, we study the switching behavior under strain. Fig. \ref{fig:piezo}(b) and (c) show switching responses observed for the same $x=0.31$ device, where (b) is the response after the device is cooled with no applied strain and (c) is the response after being cooled with strain (corresponding to an applied $40V$ to the piezoelectric cube the device was mounted on). At the current density shown, there is a change in sign of the switching response due to the applied strain. At higher current densities, the original sign is recovered (see Fig. \ref{fig:piezo_extended} in the supplement), so that there is a sign flip as a function of the pulse current density. This similarity to the $x=0.33$ sample behavior could be explained by the strain subtly altering the RKKY-dominated exchange constants and allowing a slight increase in the minority zigzag phase, as illustrated in Fig. \ref{fig:piezo} (e). Supporting this, the lattice distortion associated with magnetic ordering is small, with high-resolution synchrotron powder XRD measurements taken at low temperatures showing an $a$ lattice parameter of $5.65407 \pm 0.00015$\AA{} for the zig-zag-dominated samples and $5.65486 \pm 0.00018$\AA{}  for the stripe-dominated samples (See Figs. \ref{fig:pxrd31_300}, \ref{fig:pxrd31_4}, \ref{fig:pxrd35_10}, and \ref{fig:pxrd35_300} in the supplemental material). This corresponds to a difference on the order of $0.1\%$, which is what is achievable with the piezoelectric cube used for this experiment.  Strain can therefore tune the switching response for a $x=0.31$ device, to that of a device with $x=0.33$ -- a direct indication magneto-elastic coupling can be used to manipulate the domain structure of the magnetic texture.


The complexities of the competing order parameters notwithstanding, the collective dynamics associated with the ordered phases or with the coexisting spin-glass phase will have collective modes that can carry spin currents. Typically, scattering off conduction electrons has limited the spin decay length to nanometers, and only in a small number of insulating antiferromagnets can this be extended to microns \cite{lebrun_long-distance_2020,han_AFMmagnons,nonlocal_spin_superfluid,lebrun_transport2018}. To our knowledge, such magnons have not been used to also store information non-locally in the same antiferrromagnet. In this study we show that this unusual situation can be realized. By leveraging the intertwined order parameters of \Fex{x}, spin information is not only carried far from the regions carrying the pulse current, but can also be stored non-locally, tens of microns away from the active area.

\begin{figure}
\centering
\includegraphics[width=1\linewidth]{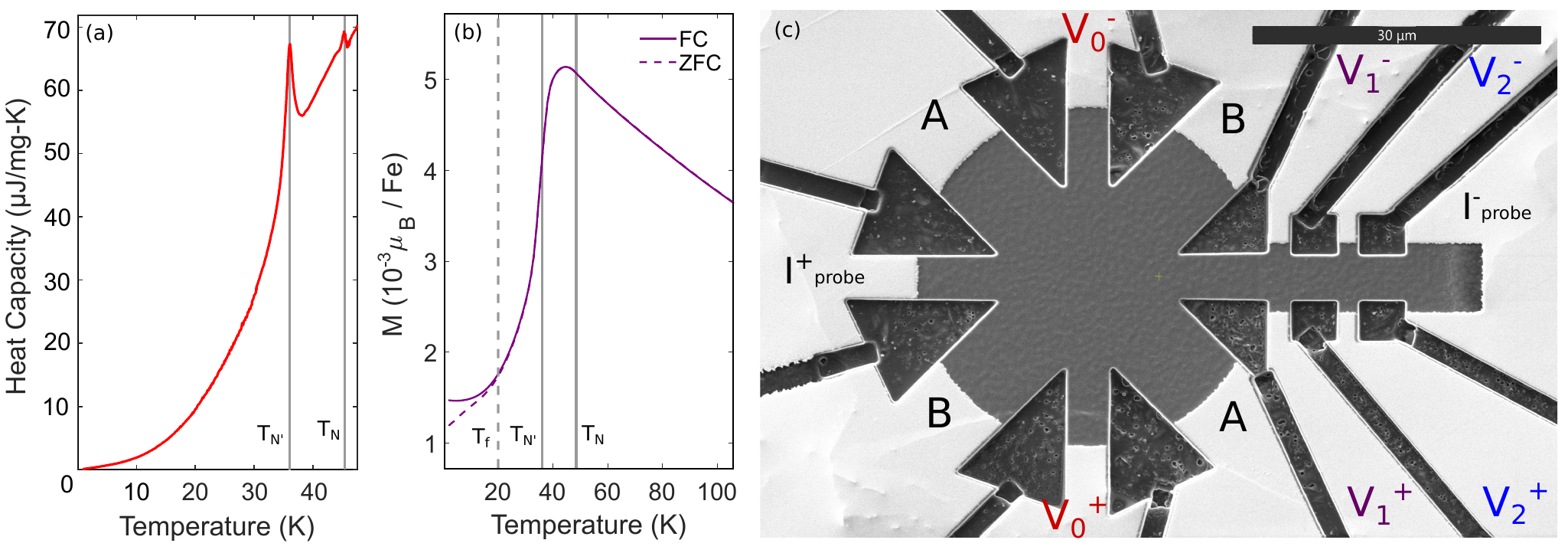}
\caption{Basic characterization of devices made of Fe$_{0.35}$NbS$_2$. (a) Heat capacity as a function of temperature. Vertical solid lines marks $T_{N}$ and $T_{N'}$, the AFM transitions. (b) Magnetization as a function of temperature measured in 1000Oe along the c-axis. The field-cooled (FC) measurement, shown as a solid curve, was measured from low to high temperature after cooling the sample in an 1000Oe field. The zero-field-cooled (ZFC) measurement, shown as a dotted curve, was measured from low to high temperature after cooling the sample with no external field. Vertical dotted and solid lines indicate the onset of the spin glass behavior ($T_f$) and the AFM transitions ($T_N$ and $T_{N'}$), respectively. (c) A switching device made from a bulk crystal. The two pulse bars are marked $A$ and $B$. The AC probe current is applied along the path marked $I_{probe}$. The local signal is measured using the contacts labeled $V_0$, and the non-local signals are measured using the contacts labeled $V_1$ and $V_2$.  }
    \label{fig:device}
\end{figure}

\section{Results}
Measurements presented in this work were primarily performed on samples of \Fex{0.35}. Heat capacity and magnetization measurements of characteristic samples are shown in Fig. \ref{fig:device} (a) and (b), respectively, showing magnetic transitions and spin glass behavior consistent with our previous characterizations of \Fex{1/3} and \Fex{0.35}.\cite{haley2020halfmagnetization, maniv_exchange_2021}  In Fig. \ref{fig:device} (c) we illustrate a device designed to measure the non-local switching response of the antiferromagnetic texture of \Fex{x}. DC current pulses are applied along the directions denoted as A and B, with a view to triggering magnons that can transport spin down the neck of the device. After the application of a pulse, the transverse resistance as measured with an MFLI lock-in amplifier using an AC probe current (denoted with $I_{probe}$) at three distinct locations goes to either a higher or lower resistance state, depending on whether an A or B pulse has been administered. More information about this can be found in the Methods section. The low temperature longitudinal resistivities of the devices measured had some small variations but were generally close to $10^{-4}  \Omega cm$. The contacts marked $V_0$, which intersect the current pulse bars, will be referred to as local, and the contacts marked $V_1$ and $V_2$ will be referred to as non-local in this paper. \\

\begin{figure}
\centering
\includegraphics[width=1\linewidth]{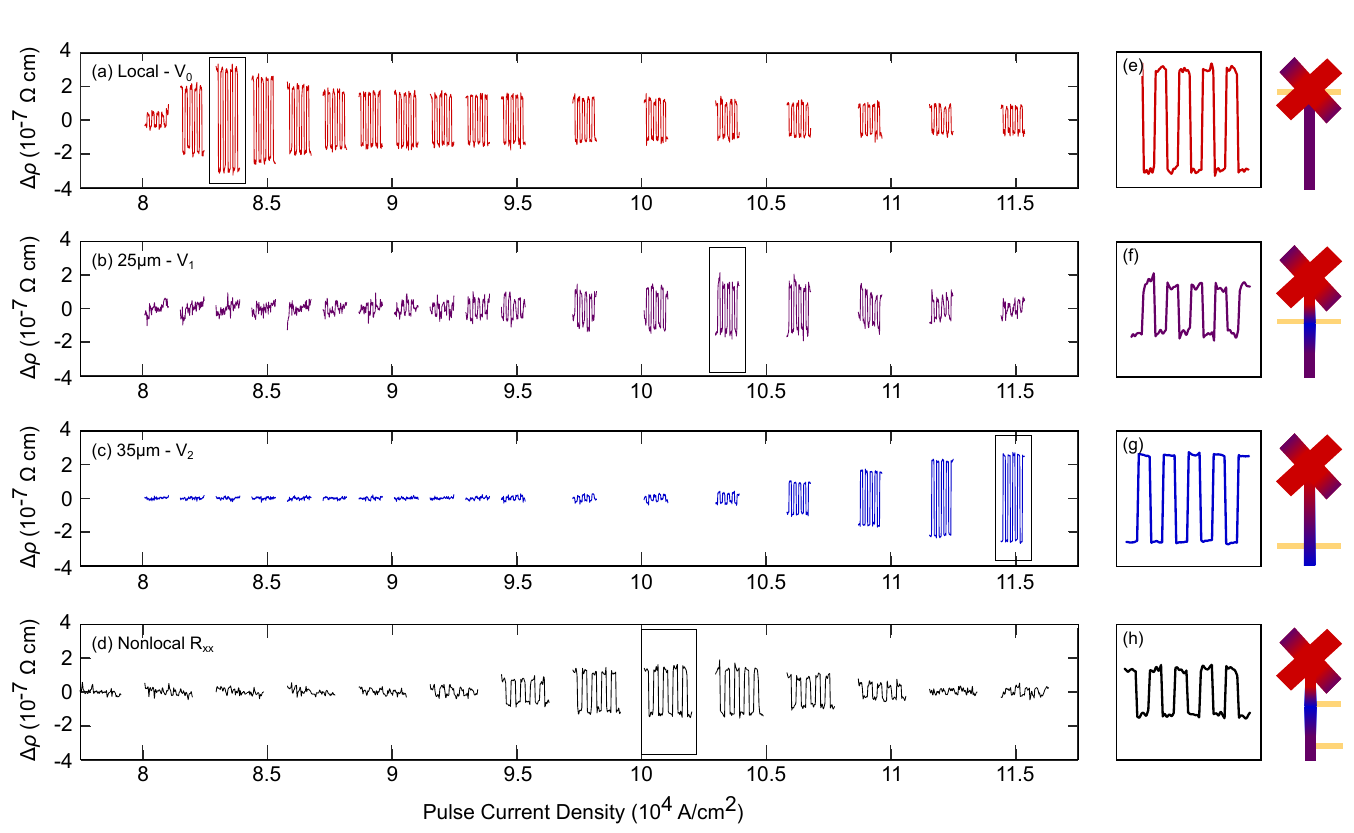}
\caption{Switching responses shown at various pulse current densities. (a) Transverse resistance response measured locally between the contacts labeled $V_0$. (c) Transverse resistance response measured $25\um$ from the center of the device, between the contacts labeled $V_1$. (e) Transverse resistance response measured $35\um$ from the center of the device, between the contacts labeled $V_2$. (d) Longitudinal resistance measured on the non-local portion of the device, between two adjacent contacts labeled $V_1$ and $V_2$. (e), (f), (g), and (h) Single sets of switching responses at the current densities indicated on the left. Schematics to the right illustrate locations of measurement contacts, with shading indicating possible domain distribution at the given current density (red and blue are perpendicular domains and purple indicates multi-domain regions).}
    \label{fig:response}
\end{figure}

Fig. \ref{fig:response} (a) shows the local response as a function of pulse current density. The response is not monotonic, instead turning on at about $8\times 10^4 A/cm^2$, quickly reaching a maximum, and then decreasing slightly to reach a stable level around $11.4\times 10^4 A/cm^2$. The measurements taken at $25\um$ and $35\um$ from the center of the active portion of the device are shown in Fig. \ref{fig:response} (b) and (c), respectively. The measurement taken $35\um$ from the center requires a larger current density to register a change from the pulses than is necessary at $25\um$ from the center, and both require larger current densities than the local response. The relative sizes of the responses vary from device to device, but the current density required is largely unchanged. Similar devices made of \Fex{x} $x\approx 1/3$ show weak local switching, but no stable switching response at the non-local contacts (supplement). The progressively larger current densities required to observe a switching response further from the active area of the device is largely consistent with the propagation of magnons, which dissipate with distance. There are two notably surprising aspects to this result, however. First, the non-local contact V$_2$, while requiring a larger current density, tends to have a larger switching response than the non-local contact V$_1$.  As discussed below, an important reason for this is that the former is closer to the edge of the crystal. Second, the non-local contacts generally exhibit an opposite switching response to the local contacts V$_0$, so that the pulse directed in the same direction (A or B) will raise the local transverse resistance and lower the non-local transverse resistance. This suggests that the preferred domain orientation upon a current pulse differs between the two regions.

\begin{figure}
\centering
\includegraphics[width=\linewidth]{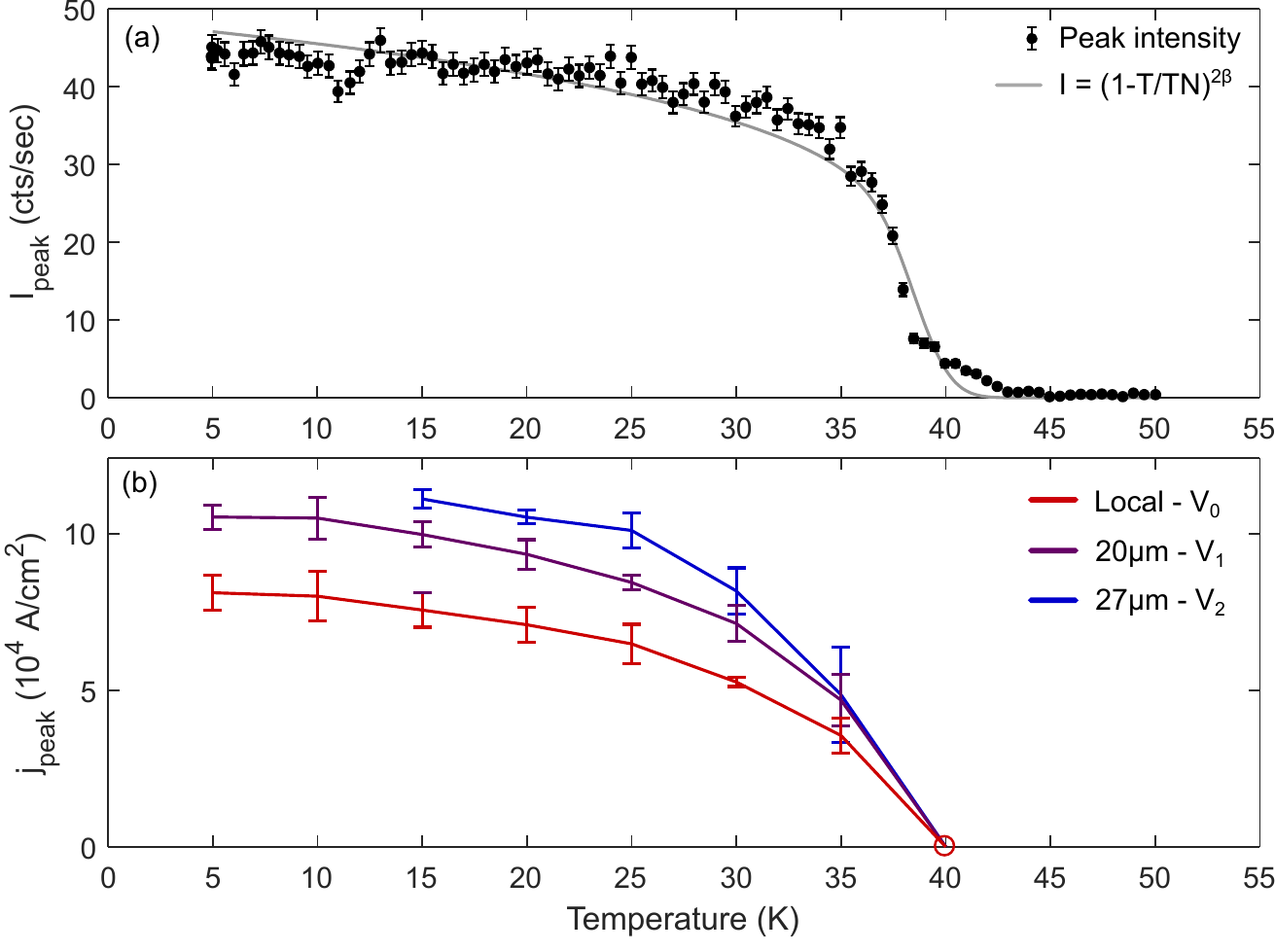}
\caption{Comparison of temperature dependencies of neutron scattering measurements and switching measurements. (a) Order parameter peak intensity measured with neutron scattering as a function of temperature, with the associated critical exponent fit, with $2\beta = 0.21(2)$. (b) Current density of peak switching response as a function of temperature, for all three sets of measurement contacts. Peak switching was determined by fitting the amplitudes of the responses to a Gaussian model, whose standard deviations give the uncertainty indicated by the error bars. The open circle at $40K$ denotes the lack of switching at this temperature regardless of current density. The full dataset can be found in the supplement.}
    \label{fig:tempdep}
\end{figure}

The response of $\rho_{xx}$ between the $25\mu m$ and $35\mu m$ non-local contact is shown in Fig. \ref{fig:response} (d). The non-local $R_{xx}$ response closely mimics the $25\mu m$ non-local $R_{xy}$ response, with a peak just below $10.5 \times 10^4 A/cm^2$. The response of $\rho_{xx}$ is notably absent where the $35\mu m$ non-local $\rho_{xy}$ response is strongest; this suggests that the full conductivity tensor is affected in the $25\mu m$ non-local region, whereas the $35 \mu m$ non-local region has a dominant response only in the off-diagonal components $\rho_{xy}$. This could be explained by increased domain-wall scattering in the former, whereas the latter has fewer domains. In order to explain the opposite response between the local and $35\mu m$ non-local switching, the average principal axis of highest conductivity must be similarly oriented in the perpendicular direction. 

Fig. \ref{fig:tempdep} (b) shows the temperature dependence of the pulse current with the maximum switching response for both local and non-local contacts; this is based on an analysis of data shown in Figs. \ref{fig:temp0}, \ref{fig:temp1}, and \ref{fig:temp2} in the supplement. These measurements were taken on a device with non-local contacts $20\mu m$ and $27\mu m$ away from the center of the active area. At all three locations on the device, the threshold switching current grows with increasing temperature below the AFM temperature, closely mimicking neutron scattering measurements of the peak intensity corresponding to the AFM order parameter (Fig. \ref{fig:tempdep}(a)). This, and the disappearance of the switching response at the N\'eel temperature, demonstrates a direct connection between the threshold current for switching and the opening of an AFM gap. 

\section{Discussion} 

The temperature dependence of the switching amplitude shown in Fig. \ref{fig:tempdep} is strongly indicative that the threshold current required for the switching is proportional to the magnitude of the antiferromagnetic order parameter. The nonmonotonic shape of the switching behavior as a function of current density observed locally is also observed in the non-local contacts, suggesting the same underlying behavior is also present in these regions. Collective excitations carry spin and rotate the spin texture in these non-local regions in the same way that the spin-polarized current pulses do in the local region.

Two unusual features from the data deserve some attention. (i) The non-local response closer to the active area has a consistently smaller signal than that closer to the crystal's edge. (ii) The second non-local region has an average principal axis of highest conductivity that is always oriented perpendicular to that of the local region. 

We suggest that both of these effects are connected by the elastic response of the system. Little {\it et al.} recently showed that the antiferromagnetic order is strongly coupled to a structural distortion. Here, we have demostrated that strain can directly control the sign of the switching response (Fig. \ref{fig:piezo}(b),(c)) \cite{little2020three}. The situation is similar to the physics of martensites, where magnetostrictive effects can prefer a multi-domain state as the total elastic energy is balanced against the energy penalty of creating a domain boundary.\cite{Gomonay2002}  For clamped samples, such as those studied here, the incompatibility between the bulk strain induced by the AFM order and the surface strain acts as an ``elastic charge'' that produces a long-range field\cite{Gomonay2002,Gomonay2002magnetism,Gomonay2007,eshelby1956} introducing a competing energy that can make multi-domain states favorable -- a situation comparable to the long-range magnetic dipole fields in FMs \cite{kittel1956}.   Therefore, away from the local regions where spin is transferred due to the driving current, the system will tend to rotate in the opposite direction in order to preserve the balance of  domains in different orientations. 

Since the clamped boundary is the originator of the long-range forces, it is natural to expect that this effect is most stark close to the crystal's edge, as seen in the $35\mu m$ non-local $\rho_{xy}$ response, while the region surrounding the closer non-local contacts would need to rearrange less. The $25 \mu m$ non-local response would then detect domain wall scattering and smaller re-orientations of the N\'eel vector, explaining its relatively smaller response in $\rho_{xy}$ and larger response in $\rho_{xx}$, which is amplified by domain boundary scattering. Finally, we note that in order for this mechanism to be effective, the internal strain of the device must be significant -- comparable to the strains applied in our experiment shown in Fig. \ref{fig:piezo}(g) of $\sim 0.1\%$ . Given that anisotropy of the lattice parameters themselves is only $\sim 0.1\%$, this suggests that the applied current pulses must orient a significant fraction of the device into a single domain, away from the active area. This also explains the reduction of $\rho_{xx}$ at higher currents, which would be lowered by the reduction of domain walls. 


Disorder \cite{kalita2001,kalita2005}, entropy \cite{kalita2001mag,Li1956}, and leakage current provide alternative explanations for the equilibrium domain configuration between pulses. Disorder-driven domain formation, however -- in which domains are tied to defects -- does not explain the stronger signal at the farther non-local contact nor its sign being opposite to the local contacts. Similarly, leakage current also does not explain why switching is so much stronger in the farther non-local contact V$_2$ than in V$_1$. Entropy-driven domain formation should be strongest close to the N\'eel temperature, which is inconsistent with the switching being enhanced as the temperature is lowered below the transition. We therefore suggest that the natural explanation is the combined action of spin-carrying collective excitations coupled to the magnetoelasticity of the system.

 Typically, metals transport spin via their conduction electrons, while magnetic insulators transport spin through collective excitations such as magnons. Conduction electron spin currents generally decay more quickly than magnon spin currents, and in practice antiferromagnetic metals in particular tend to have very short spin diffusion lengths, largely around or under $2nm$ - as is, for example, the case in Mn based alloys.\cite{ short_afm_metal,acharyya_study_2011,merodio_penetration_2014,arana_spin_2018,zhang_spin_2014}. In magnetic insulators, on the other hand, spin decay in single crystal systems has been extended to ten microns (see the case of $\alpha$-Fe$_2$O$_3$ \cite{lebrun_transport2018}). To account for the long distances of spin transport observed, the transport medium in the present system is likely to also be collective modes. This is further suggested by the particular relationship between disorder and switching that we see - while the presence of disorder should decrease the efficiency of spin transport and spin torque mediated by conduction electrons \cite{afm_spin_review}, it has been proposed as an avenue for spin superfluidity when considering the dynamics of localized electron spin \cite{ochoa2018}. We observe the strongest switching responses - both locally and non-locally - in samples with excesses or deficiencies of iron. While these are the samples with single domain types, as found through neutron scattering, they are also disordered, as can be seen in the suppression of peaks in their heat capacity, as well as in the suppression of the magnetic ordering temperature as compared to the pristine samples.

Our data suggests that the spin imparted by the current pulses is not only carried by the partially spin-polarized electrons, whose generating mechanism has been discussed in Refs.\cite{bodnar_2018, wadley_2016, zelezny, nair_electrical_2020}, but also by the collective excitations launched by the pulses, allowing regions of the sample that are tens of microns away to be switched. These scales are orders of magnitude larger than spin decay lengths of typical metallic antiferromagnets, which is a welcome discovery relevant for potential technologies based on such materials.\cite{short_afm_metal}  One question is to which order the collective excitations belong. A natural candidate is the antiferromagnetic order itself, whose magnons transfer their spin to the nearest domain wall. However, multiple pulses in the same direction would be expected to lead to incremental changes in $\rho_{xy}$ as the wall is pushed, whereas we observe single-pulse saturation of the signal. Another scenario is that there exists a collective mode capable of carrying spin that travels through domain boundaries with relative impunity. The presence of a spin glass is a candidate, arising from magnetic disorder or from the frustration of two ordered ground states.\cite{maniv_antiferromagnetic_2021} In this case, the collective modes could be the so-called Halperin-Saslow modes, theorized decades ago and recently suggested as mediums for spin transport.\cite{ochoa2018} Importantly, recent evidence suggests that the spin glass and the antiferromagnetic order are strongly exchange-coupled.\cite{maniv_exchange_2021} Future work is needed to reveal the microscopic mechanism behind the non-local response, but for now it would be interesting to see whether other electrically switchable antiferromagnets can show similar behavior. The observation of non-local switching due to collective antiferromagnetic dynamics could open a new pathway to magnonic memory and other spintronic applications of complex antiferromagnets.

\section {Methods} Single crystals of Fe$_{x}$NbS$_2$ were synthesized using a chemical vapor transport technique. A polycrystalline precursor was prepared from iron, niobium, and sulfur in the ratio $x:1:2$ (Fe:Nb:S). The resulting polycrystalline product was then placed in an evacuated quartz ampoule with iodine as a transport agent (2.2 mg/cm$^3$), and put in the hot end of a two zone MTI furnace with temperature set points of 800 and 950 for a period of 7 days. High quality hexagonal crystals with diameters up to several millimeters were obtained. \\

Devices were fabricated using the FEI Helios G4 DualBeam focused ion beam at the Molecular Foundry at LBNL. The devices were mounted on Torr Seal and sputtered with gold for electrical contact. In most cases the crystals were exfoliated to reach a thickness under $4\mu m$. The switching pulses were single square waves administered with Keithley 6221 Current Sources. \\

 Transport was measured during the switching experiments via an MFLI lock-in amplifier. An AC probe current ran through the device both during and in between switching events, and for each measurement in this work had an rms value between $25\mu A$ and $100\mu A$ and a frequency of either $277Hz$ or $1333Hz$. Measurements were also taken with the AC probe current turned off and its corresponding leads removed during the switching event itself, and the resulting switching behavior was unchanged. A range of AC probe frequencies were also tested, and aside from an increase or decrease in noise there was no measurable difference in the resulting behavior. Both of these tests can be found in the supplement.  \\

Low field magnetization measurements were performed using a Quantum Design MPMS-3 system with a maximum applied magnetic field of 7 T. AC heat capacity was measured using a Quantum Design PPMS system. \\

High-resolution wide-angle x-ray powder diffraction measurements were performed on the beamline 28-ID-1 at the National Synchrotron Light source II at Brookhaven National Laboratory. The raw data were collected by the incident beam with a wavelength of 0.1668 \AA and a Perkin-Elmer area detector, and transformed to diffraction data.  The Rietveld refinement was carried on by GSAS-II \cite{gsas}. Single-crystal neutron diffraction experiment was performed on BT-7 at the NIST center for neutron research. 

\section{Data availability} Source data are available for this paper. All other data that support the plots within this paper and other findings of this study are available from the corresponding author on reasonable request.
\section{Code availability} Code used to analyze the data in this work is available from the corresponding author on reasonable request.
\section{Acknowledgments} This work was supported as part of the Center for Novel Pathways to Quantum Coherence in Materials, an Energy Frontier Research Center funded by the U.S. Department of Energy, Office of Science, Basic Energy Sciences. Work by J.G.A. was partially supported by the Gordon and Betty Moore Foundation’s EPiQS Initiative through Grant No. GBMF9067.
\section{Author Contribution} 
S.C.H., S.T.-L., and M.A. performed crystal synthesis. S.C.H. and E.M. performed Focused Ion Beam fabrication. S.C.H. conducted switching measurements. E.M. conducted magnetization measurements. N.M. and E.M. conducted heat capacity measurements. T.C. and J.M. discussed consequences of different kinds of nonlocal magnetization switching. S.W. and R.J.B. conducted neutron scattering measurements and synchrotron x-ray powder diffraction, and analysed the resulting data. J.G.A. designed the main experiment. S.C.H. and J.G.A. performed data analysis. S.C.H., J.G.A., T.C., and J.M. wrote the manuscript with input from all coauthors. 
\section{Competing Interests} The authors J.G.A., S.H., and E.M. have submitted a patent application relevant to the contents of this manuscript. 
\section{Correspondence} Correspondence and requests for materials
should be addressed to S.H. or J.G.A.~(email: shannon\_haley@berkeley.edu, analytis@berkeley.edu).

\bibliography{EB_refs.bib}
\section{Supplementary Materials}
\subsection{Strain}

\begin{figure}
\centering
\includegraphics[width=\linewidth]{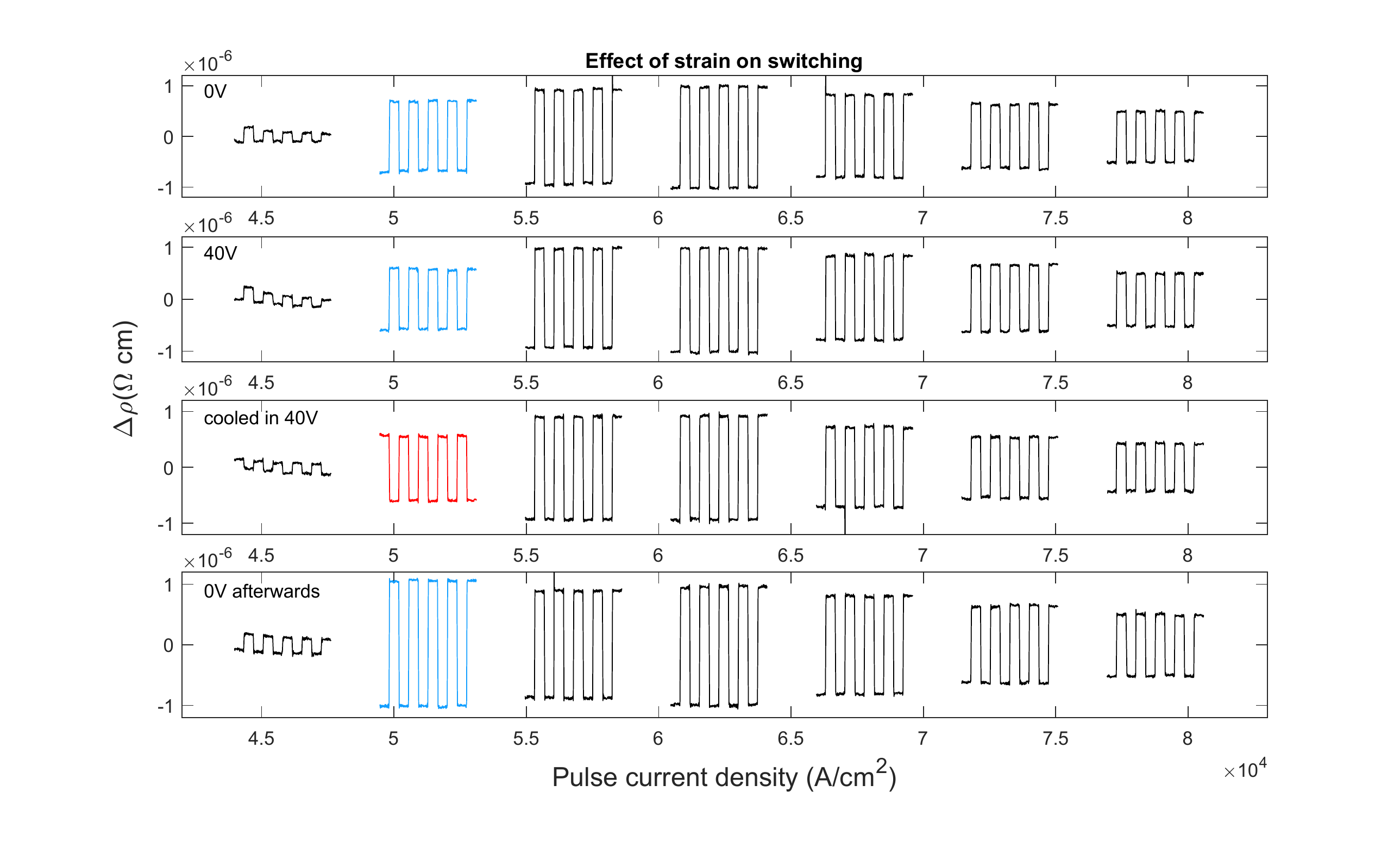}
\caption{Local switching as a function of pulse current density measured for a device with and without applied strain.}
    \label{fig:piezo_extended}
\end{figure}

Uniaxial strain measurements were performed by mounting a device on a picma chip actuator. The device was about $3.5\mu$m thick with pulse bars about $5\mu$m wide, and was mounted on Stycast. All measurements were performed at 2K. From top to bottom in Fig. \ref{fig:piezo_extended}, the switching was conducted on the device with: no voltage (strain) applied, 40V applied at 2K, 40V applied during cooldown, and no voltage applied following the previous measurements. No difference is seen when strain is applied solely at 2K, but a sign flip is observed at about $50kA/$cm$^2$ when the device is cooled with 40V applied to the piezo, highlighted in blue. The amplitude of the switching is also slightly suppressed for that preparation.
\\

\subsection{Non-local measurements in \Fex{0.33}}
Non-local measurements were performed on a sample of \Fex{0.33}. While a weak reversible switching signal is observed in the local channel, the nonlocal channel sees jumps that do not move back and forth between stable resistance states. 
\begin{figure}
\centering
\includegraphics[width=\linewidth]{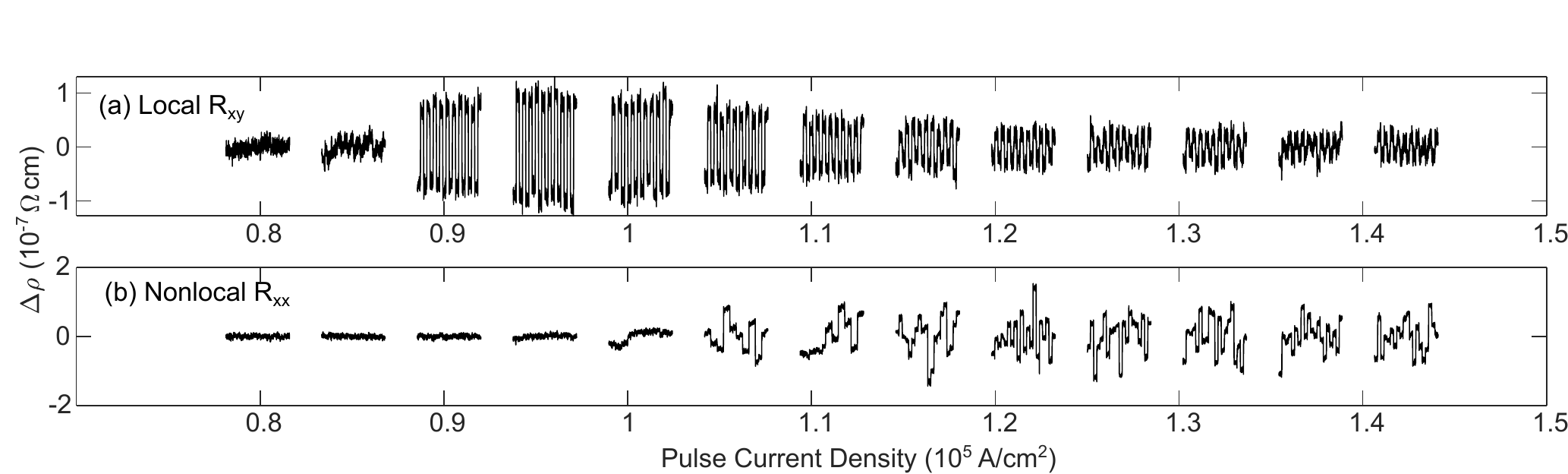}
\caption{(a) Local switching measurement as a function of pulse current density observed with a device of \Fex{0.33}. (b) Non-local longitudinal resistivity changes observed in the same device during the same switching events. }
    \label{fig:Fe33}
\end{figure}

\subsection{PXRD}
\begin{figure}
\centering
\includegraphics[width=\linewidth]{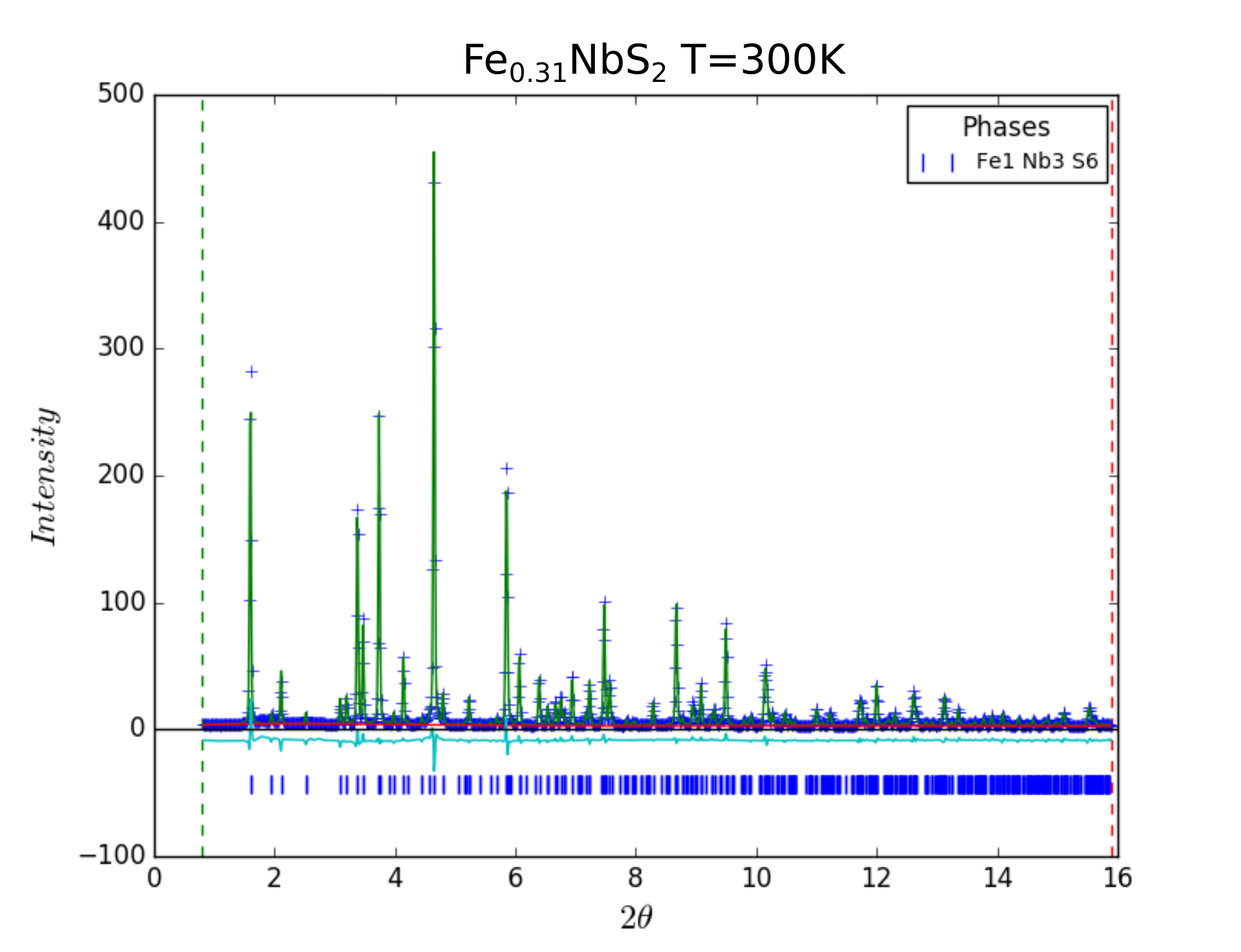}
\caption{Rietveld refinement of high-resolution synchrotron powder XRD measurements on \Fex{0.31} at $300K$. Calculated lattice parameters are   a = 5.662078  +/- 0.000137 \AA,  c = 11.964578  +/-  0.000302 \AA. The cross markers are data with the fit shown by the green curve, and the difference between the fit and data is shown in cyan. Vertical lines denote structural peak positions.}
    \label{fig:pxrd31_300}
\end{figure}

\begin{figure}
\centering
\includegraphics[width=\linewidth]{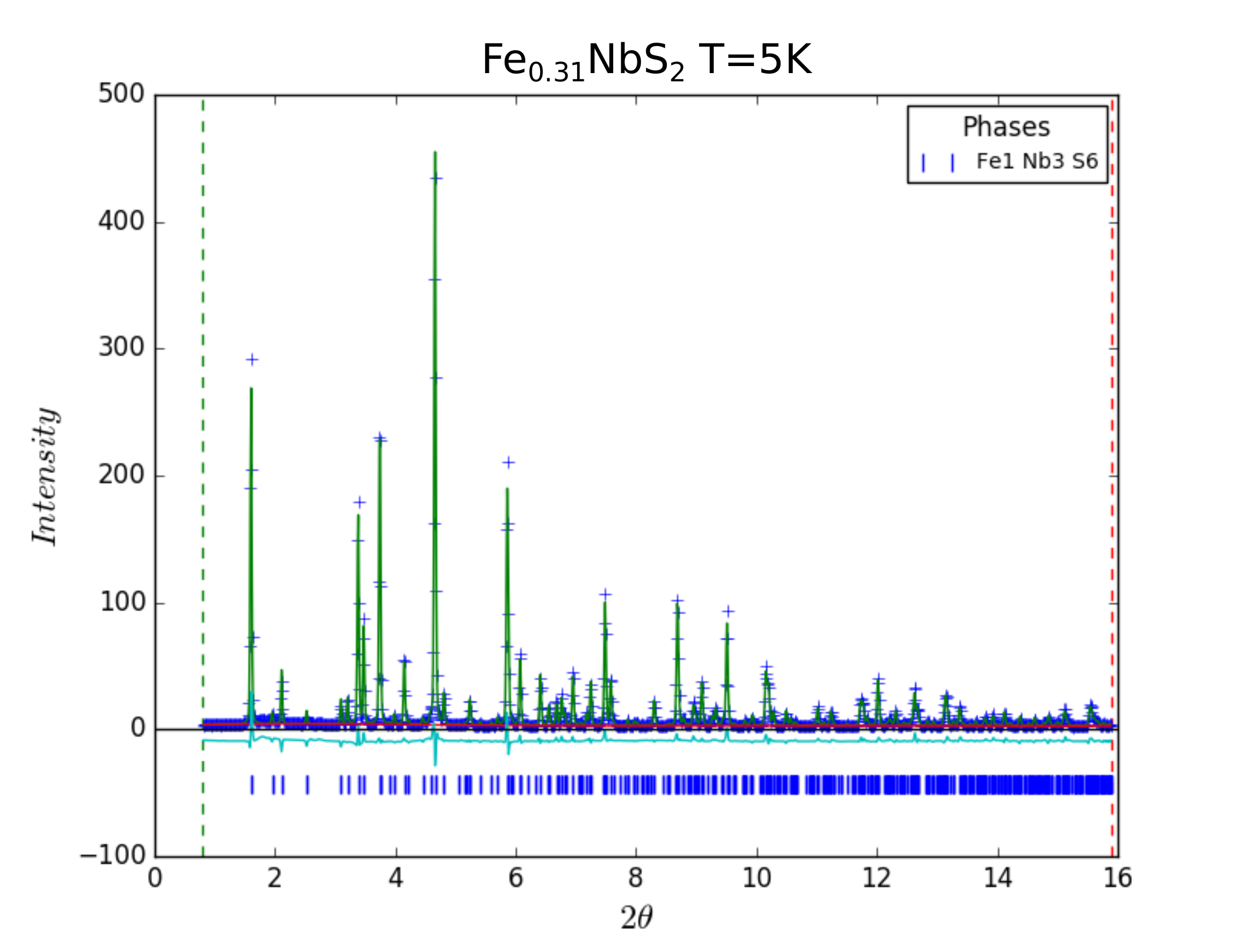}
\caption{Rietveld refinement of high-resolution synchrotron powder XRD measurements on \Fex{0.31} at $5K$. Calculated lattice parameters are   a= 5.654858 +/- 0.000182 \AA,    c=11.935061 +/- 0.000401 \AA. The cross markers are data with the fit shown by the green curve, and the difference between the fit and data is shown in cyan. Vertical lines denote structural peak positions.}
    \label{fig:pxrd31_4}
\end{figure}

\begin{figure}
\centering
\includegraphics[width=\linewidth]{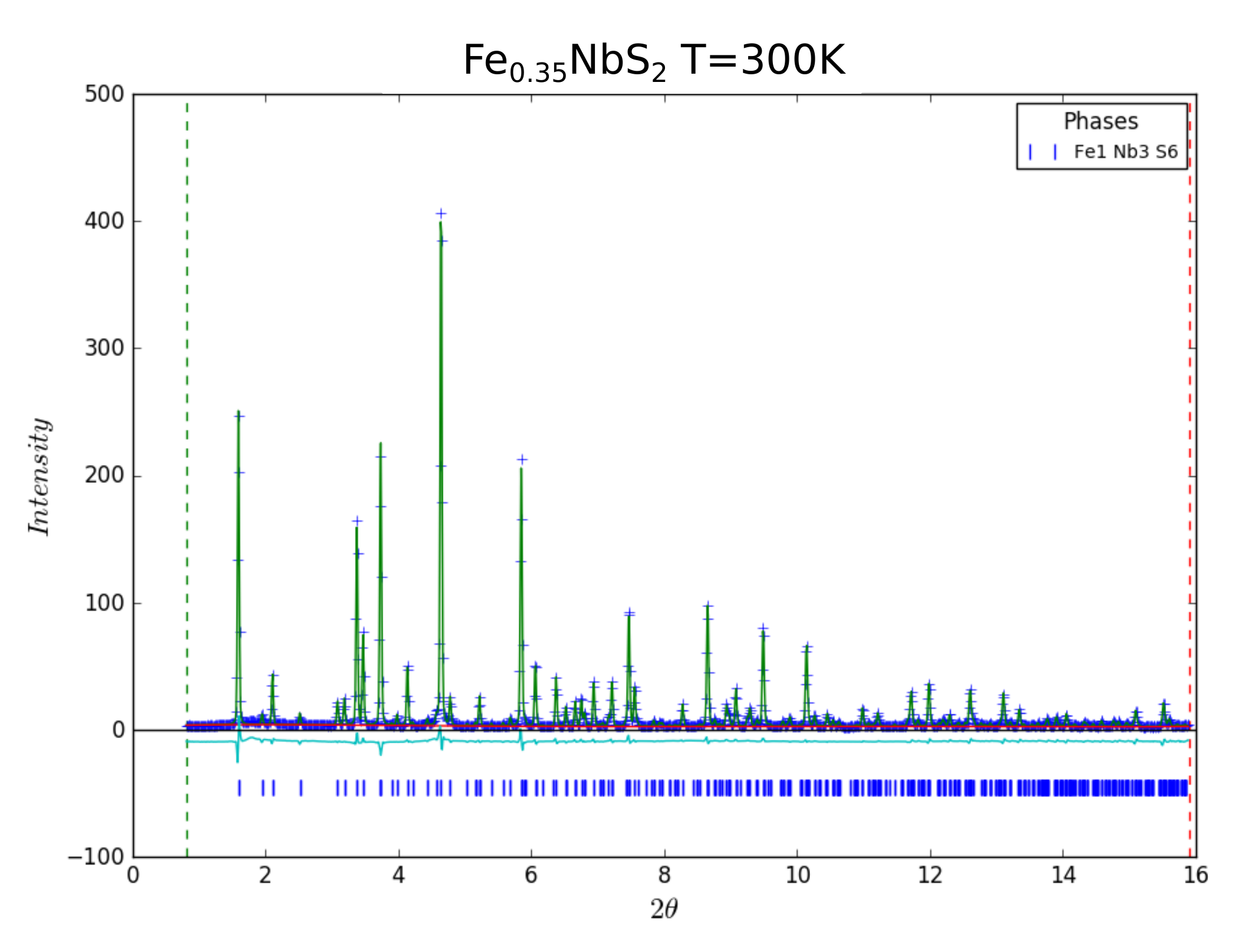}
\caption{Rietveld refinement of high-resolution synchrotron powder XRD measurements on \Fex{0.35} at $300K$. Calculated lattice parameters are  a=5.660797 +/- 0.000115 \AA,   c = 11.994610 +/-0.000260 \AA. The cross markers are data with the fit shown by the green curve, and the difference between the fit and data is shown in cyan. Vertical lines denote structural peak positions.}
    \label{fig:pxrd35_300}
\end{figure}

\begin{figure}
\centering
\includegraphics[width=\linewidth]{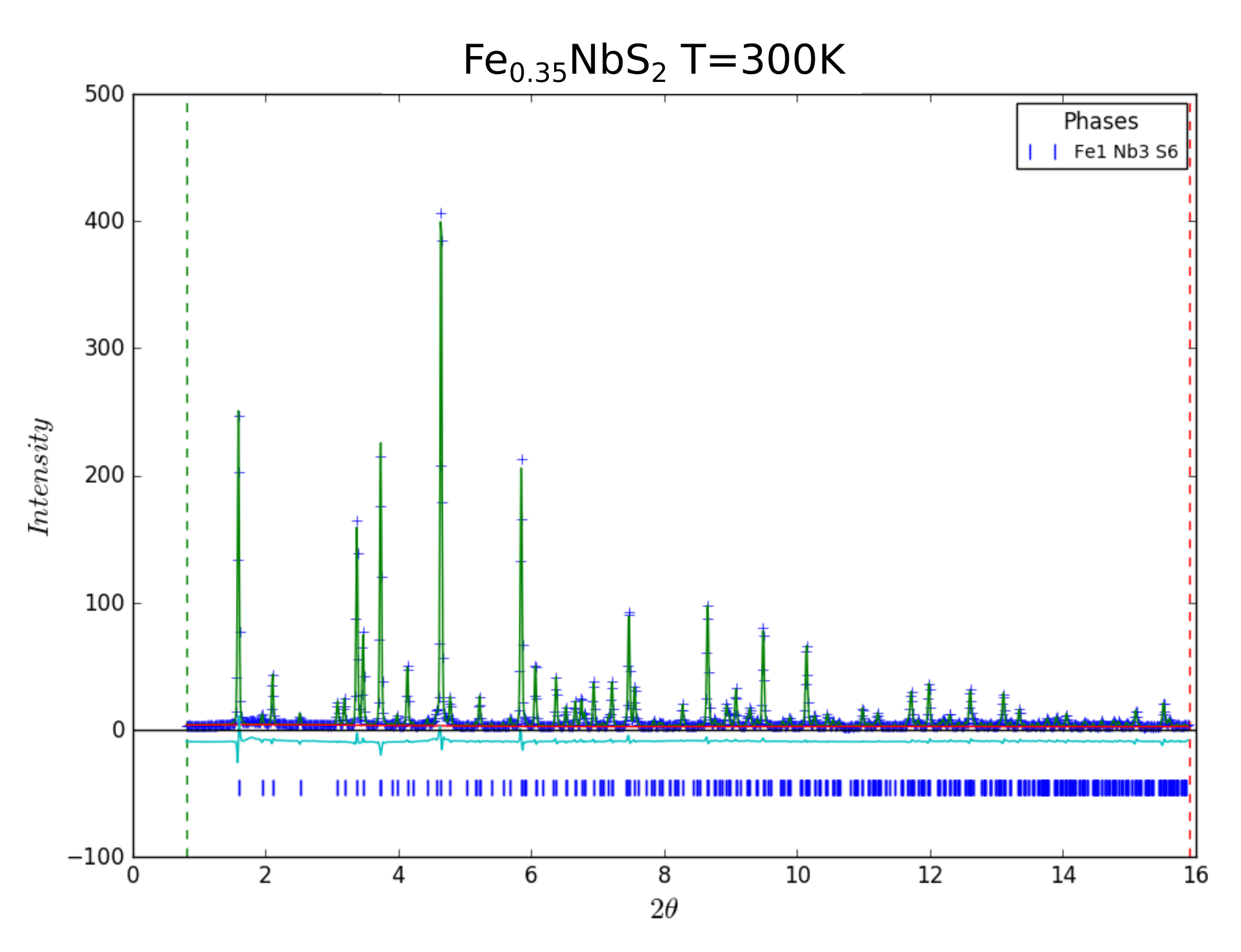}
\caption{Rietveld refinement of high-resolution synchrotron powder XRD measurements on \Fex{0.35} at $10K$. Calculated lattice parameters are  a=5.654070 +/- 0.000153 \AA,  c = 11.968898 +/-0.000296 \AA. The cross markers are data with the fit shown by the green curve, and the difference between the fit and data is shown in cyan. Vertical lines denote structural peak positions.}
    \label{fig:pxrd35_10}
\end{figure}

High-resolution synchrotron powder x-ray diffraction measurements were taken on samples of \Fex{0.31} and \Fex{0.35} at room and low temperature, as shown in Figs. \ref{fig:pxrd31_300}, \ref{fig:pxrd31_4}, \ref{fig:pxrd35_300}, and \ref{fig:pxrd35_10}. Rietveld refinements were performed to determine the lattice paramters in each of these cases.  

\subsection{Full temperature dependence}
\begin{figure}
\centering
\includegraphics[width=\linewidth]{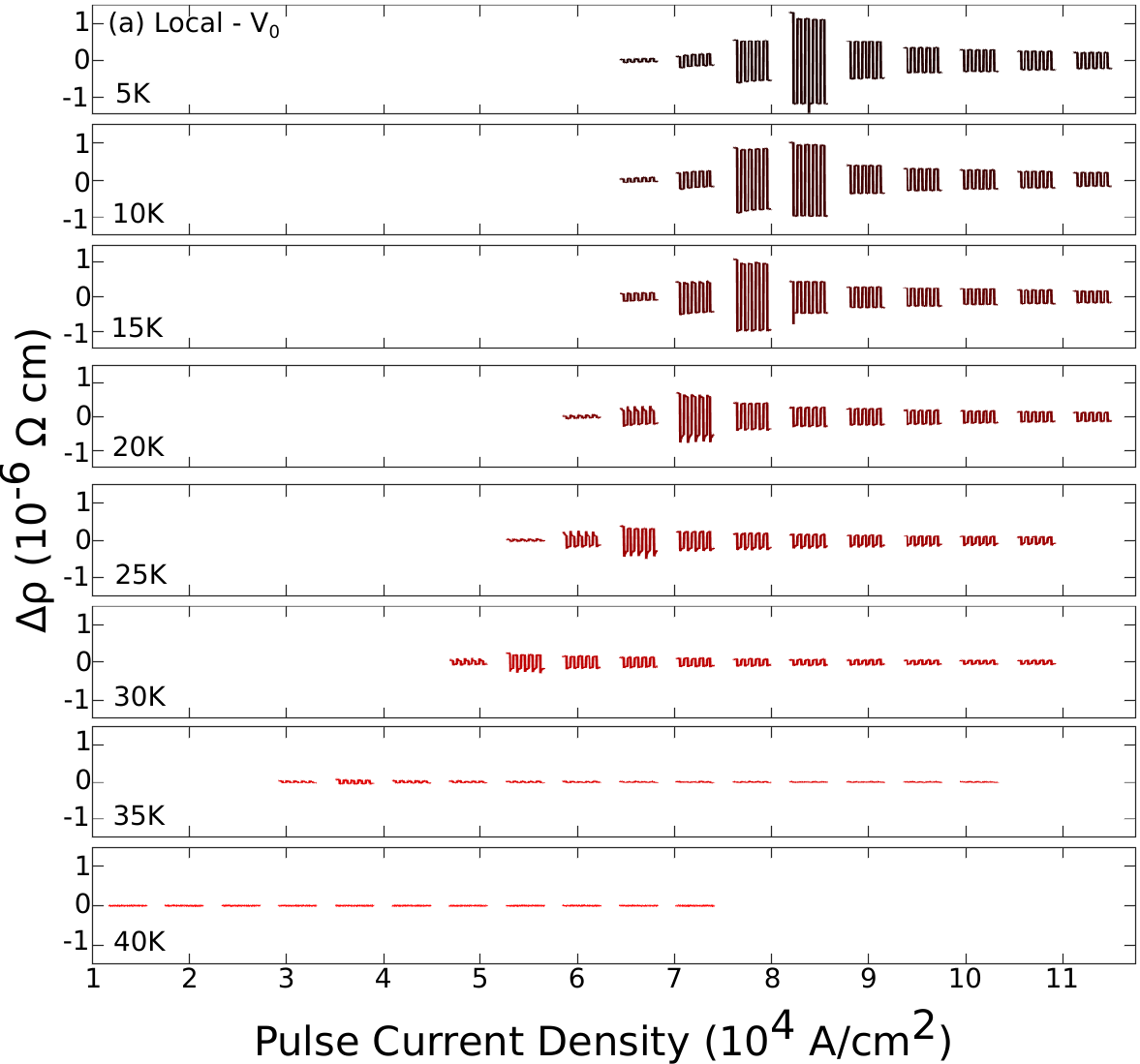}
\caption{Local switching measurement as a function of pulse current density, shown at temperatures from $5K$ to $40K$.}
    \label{fig:temp0}
\end{figure}

\begin{figure}
\centering
\includegraphics[width=\linewidth]{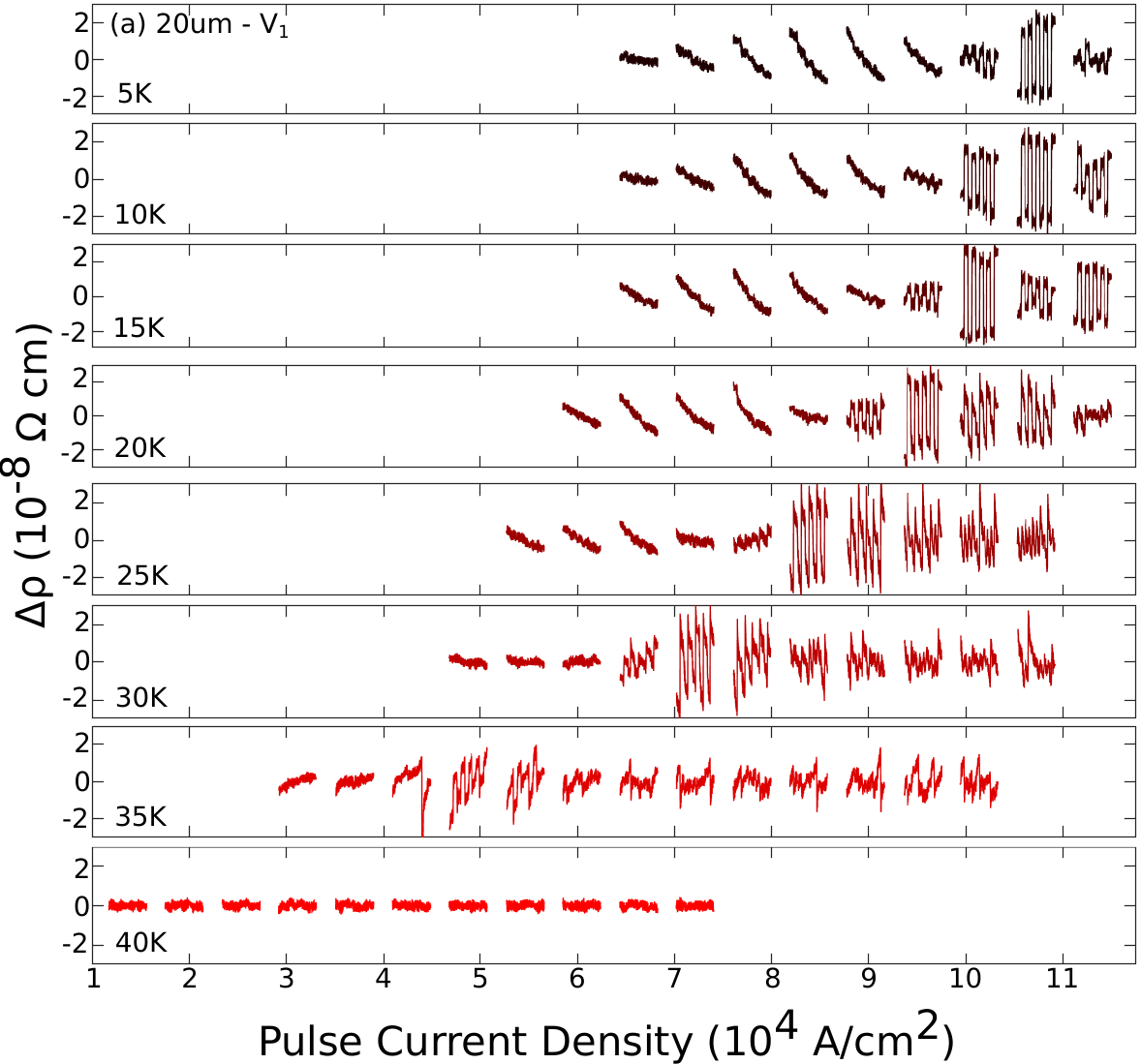}
\caption{non-local ($20\um$ from the center of the device) switching measurement as a function of pulse current density, shown at temperatures from $5K$ to $40K$.}
    \label{fig:temp1}
\end{figure}

\begin{figure}
\centering
\includegraphics[width=\linewidth]{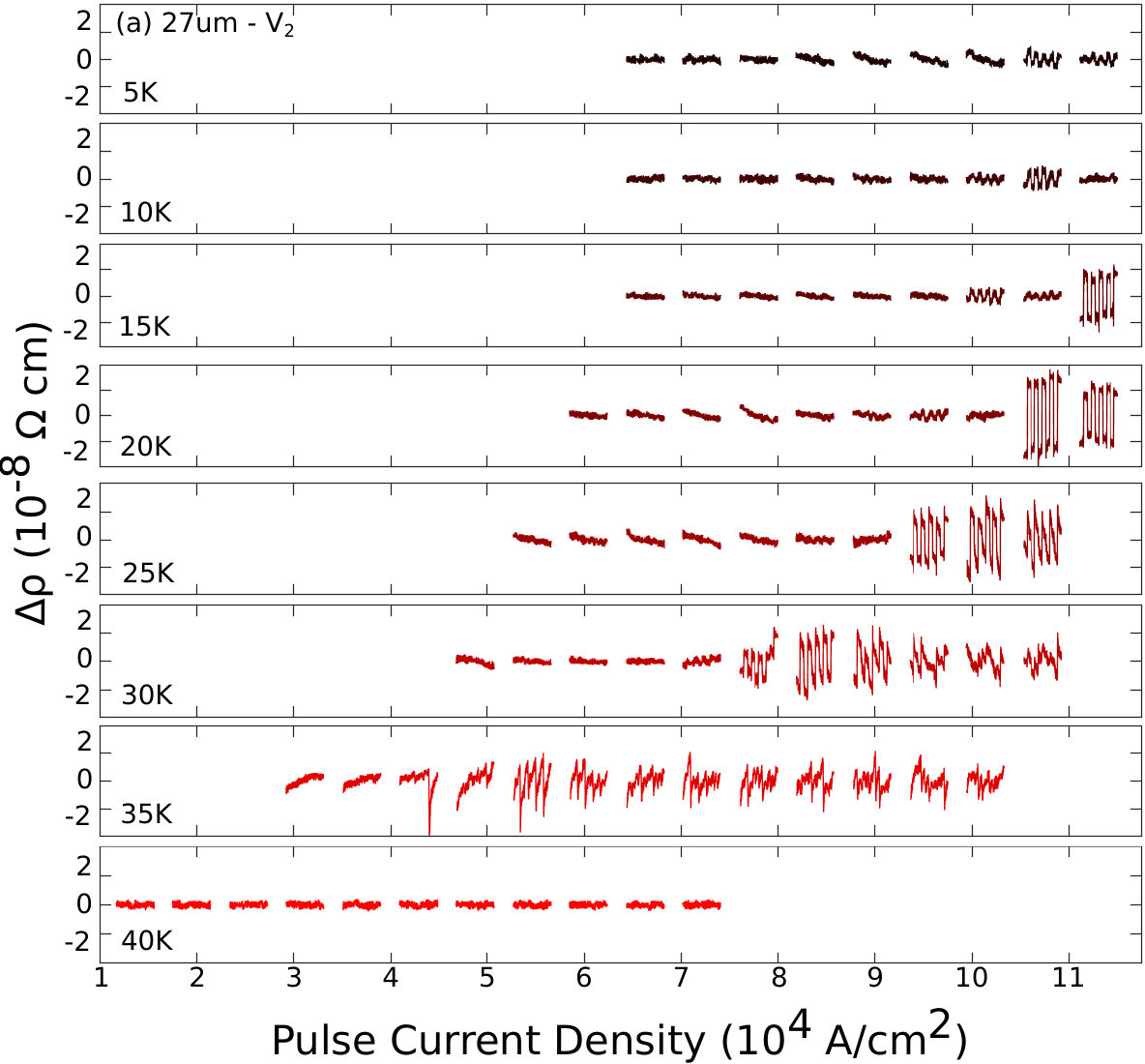}
\caption{non-local ($27\um$ from the center of the device) switching measurement as a function of pulse current density, shown at temperatures from $5K$ to $40K$.}
    \label{fig:temp2}
\end{figure}

Current density dependence was measured every $5K$ between $5$ and $40K$ for local and non-local contacts. These pulse trains are shown in Figs.\ref{fig:temp0}, \ref{fig:temp1}, and \ref{fig:temp2}. 

\subsection{Switching with lead detachment}
\begin{figure}
    \centering
    \includegraphics[width=\linewidth]{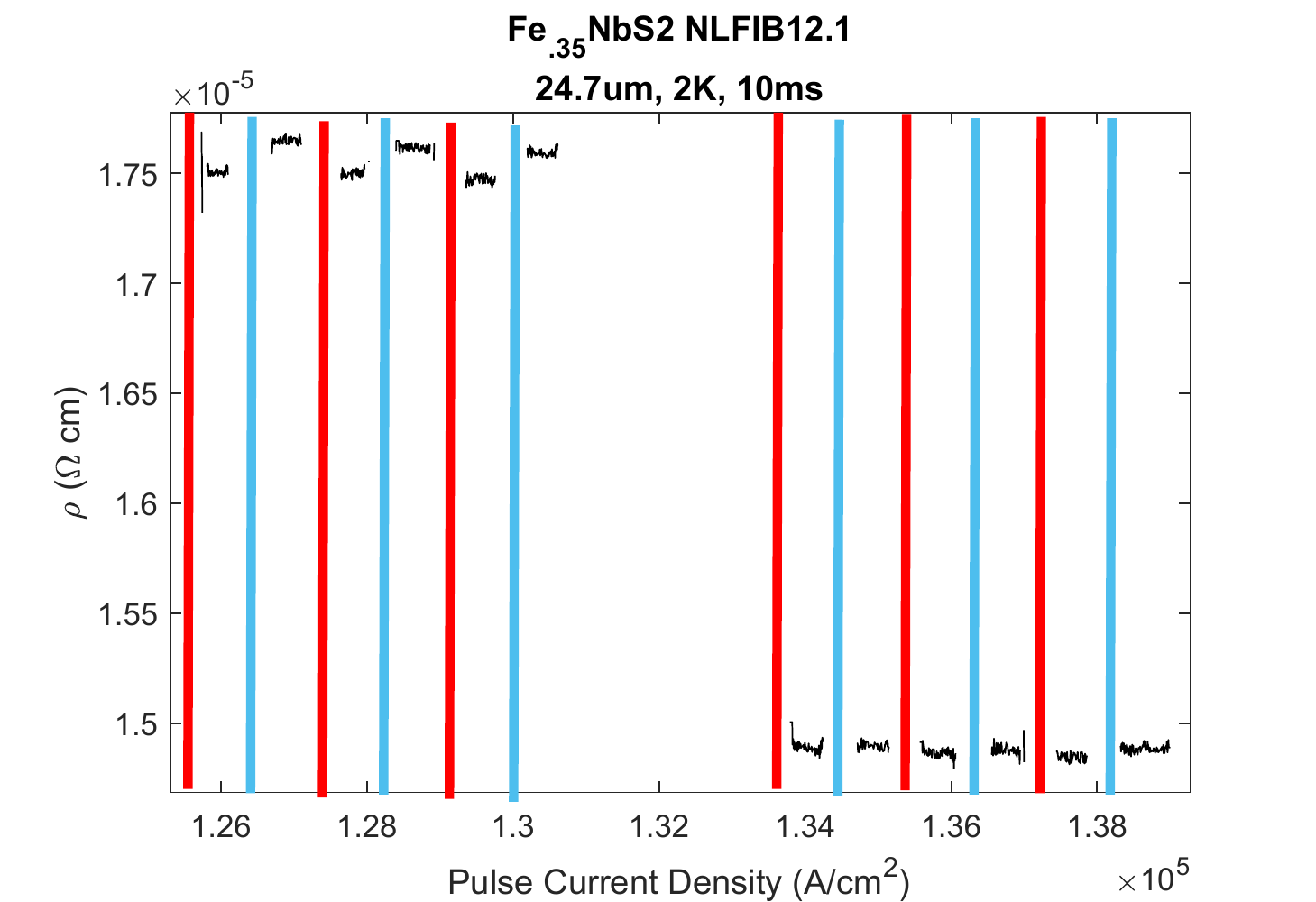}
    \caption{Switching measured $25\um$ from the center of a device at a current density that results in switching (left) and that does not result in switching (right), with the AC probe current turned off and its leads detatched during the switching events themselves. Vertical lines indicate switching events, and readouts from the lockins while the leads were detached have been omitted. }
    \label{fig:detatch25}
\end{figure}

\begin{figure}
    \centering
    \includegraphics[width=\linewidth]{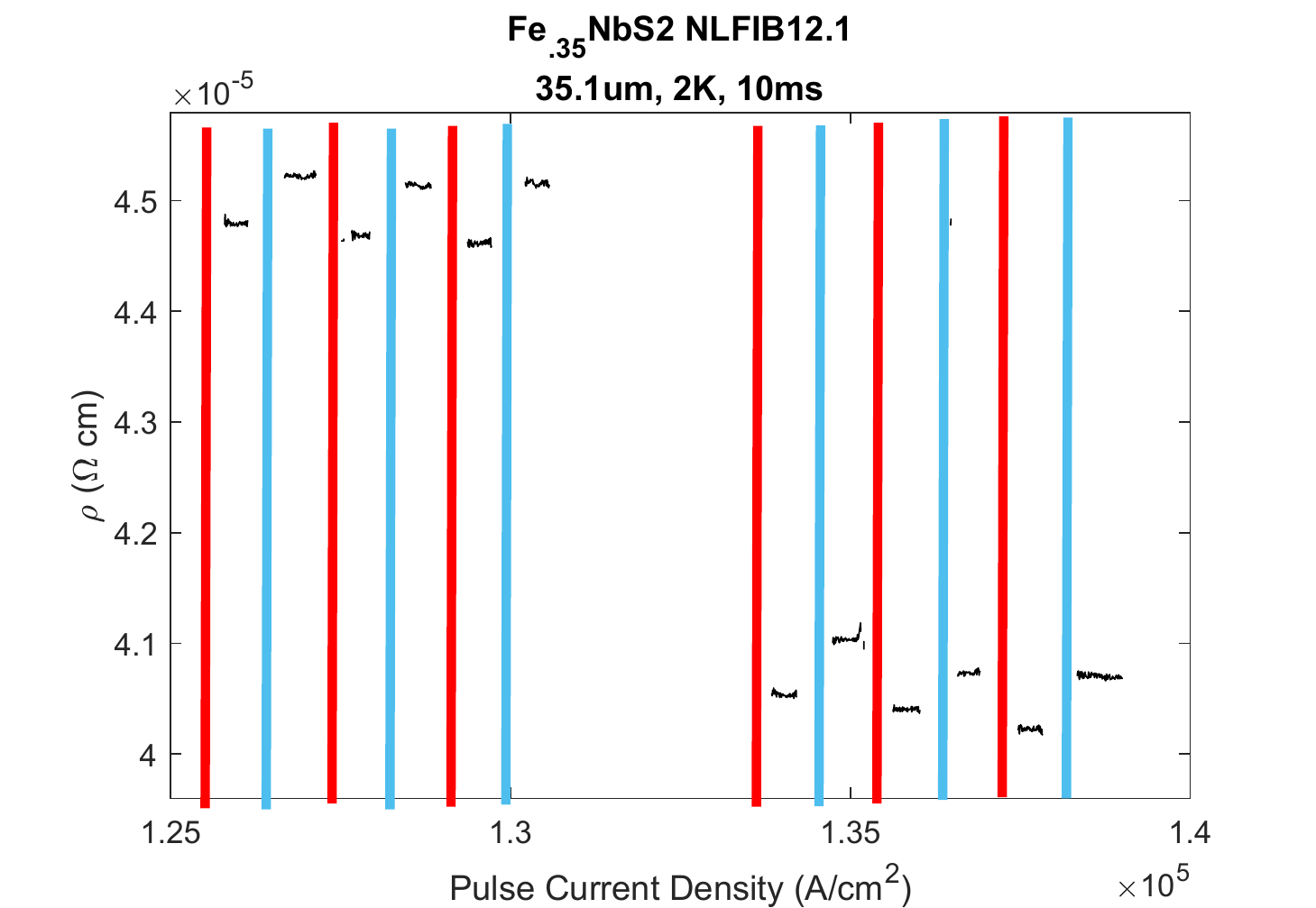}
    \caption{Switching measured $35\um$ from the center of a device at two different current densities that result in switching (left and right), with the AC probe current turned off and its leads detatched during the switching events themselves. Vertical lines indicate switching events, and readouts from the lockins while the leads were detached have been omitted.}
    \label{fig:detatch35}
\end{figure}

Measurements were performed wherein the AC probe current was turned off and its leads were detached during the switch events. The leads were reattached and the probe current turned back on in order to measure the resistance between switching events. This is shown in Figs. \ref{fig:detatch25} and \ref{fig:detatch35}. The non-local switching behavior persists, and its character does not change from that seen when the AC probe current is always on. 

\subsection{Probe current frequency dependence}
\begin{figure}
    \centering
    \includegraphics[width=\linewidth]{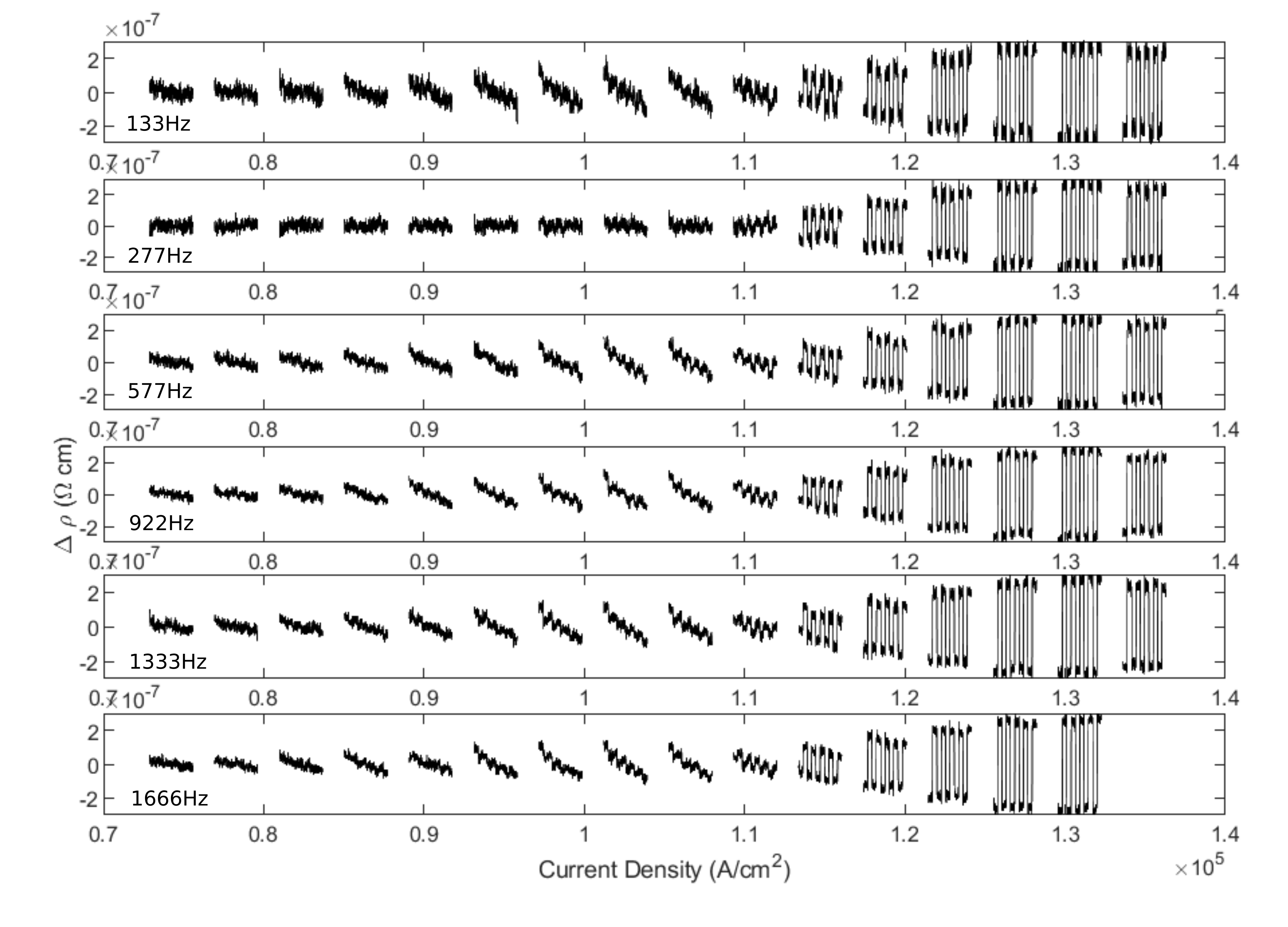}
    \caption{Switching measured $35\um$ from the center of a device at a current density that result in switching, with AC probe frequencies ranging from $133Hz$ to $1666Hz$. The noise changes from measurement to measurement, but the switching behavior notably does not. }
    \label{fig:freq}
\end{figure}

The non-local switching behavior has no discernible dependence on the frequency of the AC probe current. See Fig. \ref{fig:freq}.

\subsection{Schematic}
\begin{figure}
\centering
\includegraphics[width=\linewidth]{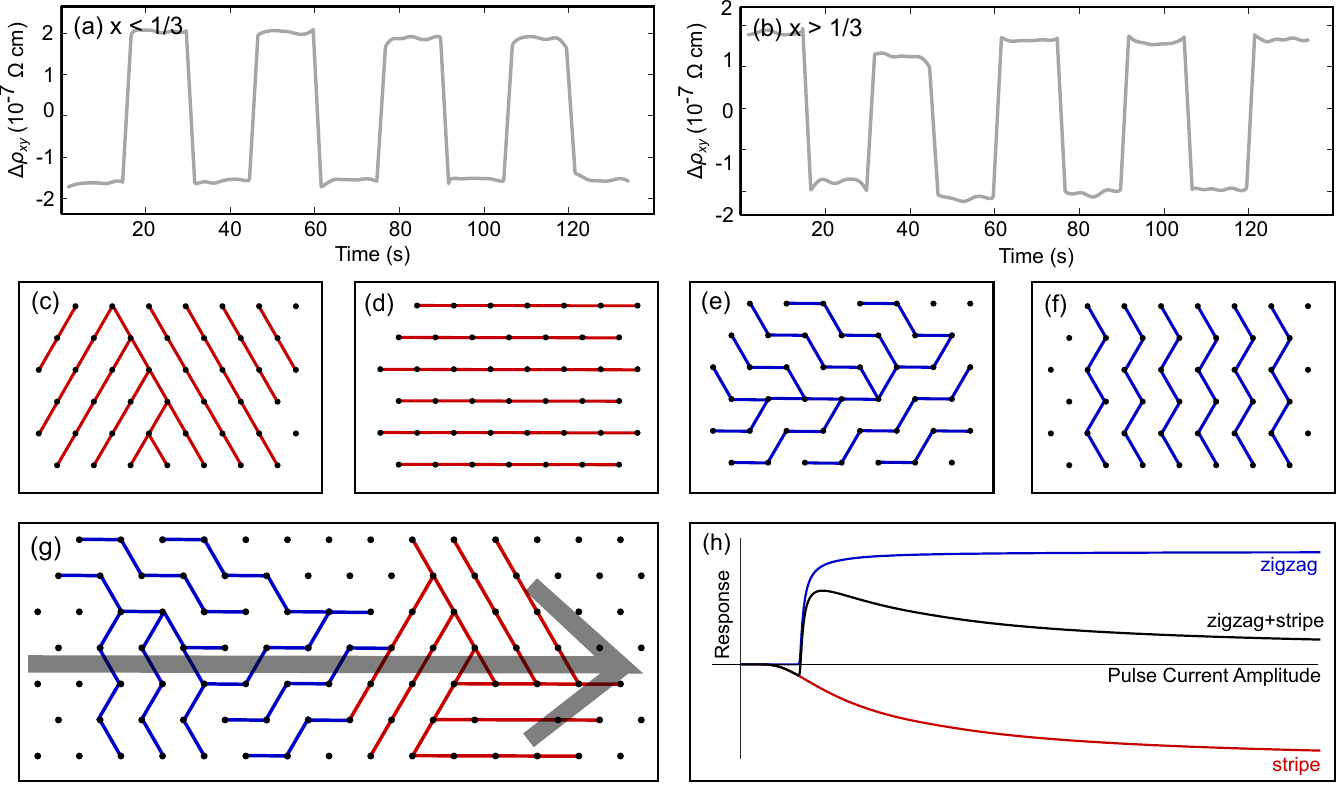}
\caption{(a) Transverse resistance switching response in \Fex{x} with $x<1/3$. Data were taken at $2K$, with pulse current amplitudes of approximately $15\times 10^4 A/cm^2$, in the regime where switching has moved beyond the initial anomalous region as seen around $8.5\times 10^4 A/cm^2$ in Fig.\ref{fig:response}A. In both cases, the first pulse and then every other subsequent pulse was normal to a crystal facet.   (b) Transverse resistance switching response in \Fex{x} with $x>1/3$. With identical device geometries, a pulse which brought $x<1/3$ to a low resistance state brings $x>1/3$ to a high resistance state, and vice versa. (c-f) Illustration of stripe and zigzag domains. Circles are iron atoms in one plane. Lines drawn between iron atoms indicate their spins are aligned. (c) Domain configuration preferred following a horizontal pulse in a stripe-dominated sample. (d) Domain configuration preferred following a vertical pulse in a stripe-dominated sample. (e) Domain configuration preferred following horizontal pulse in a zigzag-dominated sample. (f) Domain ocnfiguration preferred following vertical pulse in a zigzag-dominated sample. (h) Proposed combination of zigzag and stripe responses in zigzag-dominated sample. Note the similarity between the black curve and the observed signal in Fig. \ref{fig:tempdep}, with a small intitial response with an opposite sign flip to the main response, and a decreasing response after an initial peak. }
    \label{fig:szz}
\end{figure}

A schematic view of the proposed mechanism for resistance switching is shown in Fig.\ref{fig:szz}. Following a horizontal current pulse in a stripe-dominated sample, the domain with principal axis parallel to the current pulse is disfavored, resulting in a combination of the other two domain orientations, while the domain with principal axis perpendicular to the pulse is favored following a vertical current pulse. In a zigzag-dominated sample as well, current pulses favor domains whose principal axes are not parallel to the pulse. The domain configurations in panels (c) and (e) have opposite conductivity anisotropies, as do those in panels (d) and (f) \cite{weber2021}, so that (g) when stripe and zig-zag orders coexist, there will be competing switching responses, as shown schematically in (h). Note the similarity between the black curve in (h) and the observed signal in Fig. \ref{fig:tempdep}, with a small intitial response with an opposite sign flip to the main response, and a decreasing response after an initial peak.

\subsection{More devices}
\begin{figure}
\centering
\includegraphics[width=\linewidth]{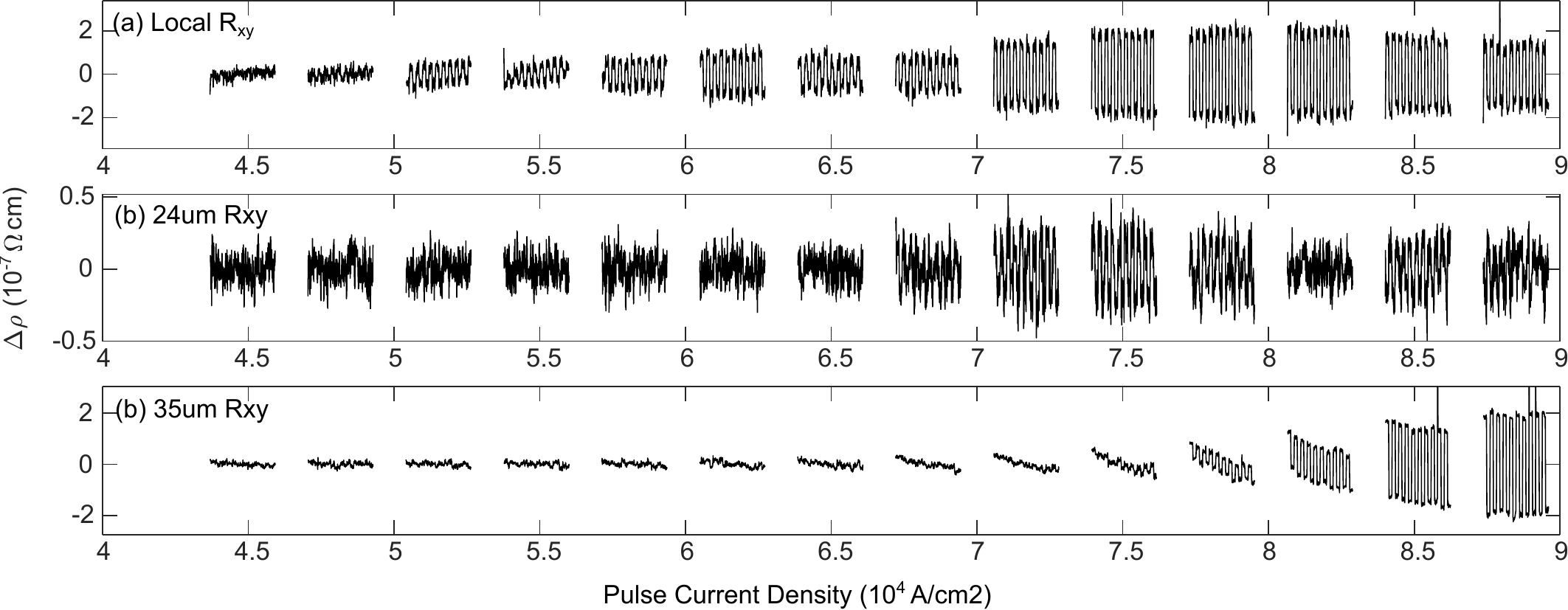}
\caption{Switching response in a device with a different current density dependence than those shown in the main text. (a) Local response is nonmonotonic, showing a change in sign as a function of pulse current density. (b) $24\um$ from the center of the device, the non-local response is small but shows a change in sign as well. The first switching responses have an opposite sign to the first local switching responses. (c) $35\um$ from the center of the device, the non-local response is comparable to the maximum local response. Compared to the initial onset of switching in the local portion of the device, the response seen here is larger and has an opposite sign. }
    \label{fig:NL8}
\end{figure}

\begin{figure}
\centering
\includegraphics[width=\linewidth]{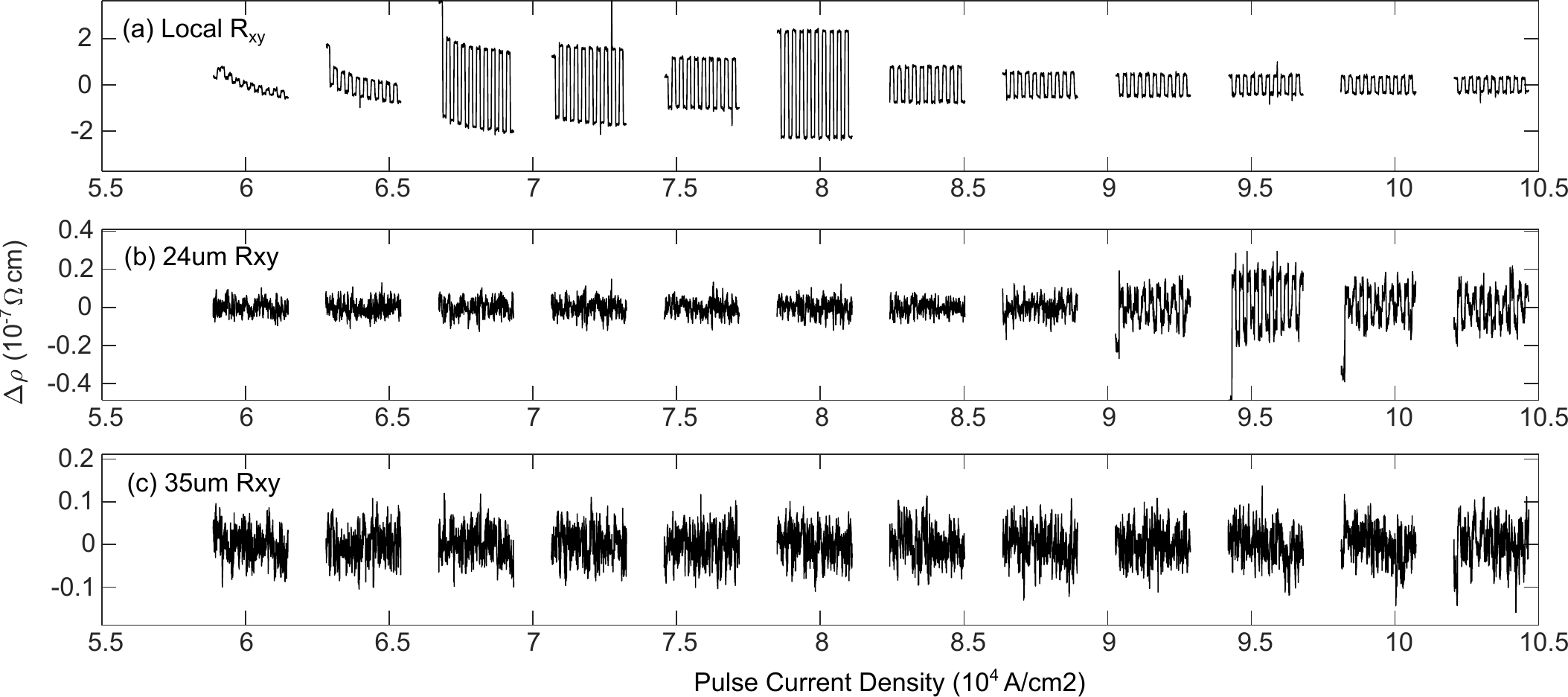}
\caption{Switching response in a device with a different current density dependence than those shown in the main text. (a) Local response is nonmonotonic, showing two peaks followed by the decreasing amplitude characteristic of most devices. The presence of two peaks is suggestive of inhomogeneous iron content or a twist in the stack of layers. (b) $24\um$ from the center of the device, the non-local response is small but has a sign opposite that of the local response. (c) $35\um$ from the center of the device, a non-local response is not observed, presumably because the measurement did not extend to high enough current densities.  }
    \label{fig:NL10}
\end{figure}

There is some variation in switching response between devices, as their dimensions, exact concentrations, geometries, and mounting conditions vary slightly. See Figs. \ref{fig:NL8} and \ref{fig:NL10} for examples. 

\clearpage


\end{document}